# Plasmonically enhanced graphene photodetector featuring 100 GBd, high-responsivity and compact size

*Ping Ma,*[\*,1,†] *Yannick Salamin,*[\*, 1,†] *Benedikt Baeuerle*[1], *Arne Josten*[1], *Wolfgang Heni*[1], *Yuriy Fedoryshyn*[1], *Alexandros Emboras*[1] *and Juerg Leuthold*[\*,1]

[1]ETH Zurich, Institute of Electromagnetic Fields (IEF), 8092 Zurich, Switzerland

†These authors contributed equally to this work.

*Corresponding authors:
ping.ma@ief.ee.ethz.ch, yannick.salamin@ief.ee.ethz.ch, juerg.leuthold@ief.ee.ethz.ch

ABSTRACT. Graphene has shown great potentials for high-speed photodetection. Yet, the responsivities of graphene-based high-speed photodetectors are commonly limited by the weak effective absorption of atomically thin graphene. Here, we propose and experimentally demonstrate a plasmonically enhanced waveguide-integrated graphene photodetector. The device which combines a 6 μm long layer of graphene with field-enhancing nano-sized metallic structures, demonstrates a high external responsivity of 0.5 A/W and a fast photoresponse way beyond 110 GHz. The high efficiency and fast response of the device enables for the first time 100 Gbit/s PAM-2 and 100 Gbit/s PAM-4 data reception with a graphene based device. The results show the potential of graphene as a new technology for highest-speed communication applications.

## Introduction

Graphene is a highly functional two-dimensional (2D) material which has enabled a plethora of new applications and devices ranging from electronics, bioengineering, and photonics. In particular, optoelectronic devices, such as high-speed photodetectors[1], take full advantage of graphene's superior electrical and optical characteristics[2], including the ultrafast carrier dynamics[3,4], high carrier mobility[5], broad spectral photoresponse[6-10], and photocarrier multiplication[11,12]. High-speed photodetectors are crucial components for many applications including high-speed optical communications[6,13], microwave photonics[14], and terahertz (THz) technologies[15,16]. Of special interest for next generation 100 Gbit/s optical communications[17] would be a graphene photodetector featuring simultaneously a high responsivity, high-speed capability and a compact footprint that can be fabricated on an integrated universal platform such as silicon (Si) photonics[6,13].

High-speed operation up to a few hundred Gigahertz has been predicted for graphene photodetectors[3,18]. To date, graphene photodetectors have already demonstrated successful data reception up to 50 Gbit/s two-level pulse-amplitude-modulation (PAM-2) signals[19] and 100 Gbit/s four-level PAM-4 encoded signals[20]. Also, very recently a 110 GHz bandwidth[21] and a bandwidth beyond 67 GHz with indications for a 128 GHz bandwidth based on power meter measurements[22] have been reported, respectively. Graphene is thus foreseeably in competition with established III/V and Ge high-speed photodetector technologies[23-25]. Yet, most graphene-based high-speed photodetectors suffer from weak external responsivities. These have their origin in the weak light-matter interaction with the atomically thin graphene and consequently a small optical absorption, which results in a relatively low external responsivity. To enhance the light interaction with graphene, a variety of solutions have been proposed. For instance, hybrid sensitized graphene photodetectors have been explored by introducing strong light absorbers such as quantum dots[26]. However, these devices have limited rise times on the order of 10 milliseconds, as they require slow carriers for an efficient multiplication process. Alternatively, graphene can be combined with optical structures such as resonant cavities[27,28] and planar optical waveguides[6,19,29-33]. While the latter is potentially the most promising approach for the integration with Si photonics, the mode overlap with graphene in a waveguide-integrated photodetector is still limited. Typical devices need to be a few tens of micrometers long to effectively absorb the evanescently coupled incident light, yielding external responsivities of a few 10 mA/W and are reaching 0.36 A/W with clever device designs[21,29,33]. In order to further improve the efficiency and scale down the length of the device to a few micrometers, detectors can be enhanced by plasmonics[34,35]. Plasmonics offers strong sub-wavelength mode confinement and local field enhancement. Quite a few plasmonic graphene concepts have already been introduced[9,15,16,31,36-43], reaching 0.6 A/W responsivity at visible wavelengths for top illuminated structure[9]. However, demonstrating a waveguide-integrated graphene photodetector in the telecom wavelengths with highest responsivity at highest speed (e.g. for a 100 GBd application) in a low footprint design is a challenge to this day.

In this letter, we present a compact, highly efficient and high-speed plasmonically enhanced waveguide-integrated photodetector comprising of a graphene layer. The photodetector is co-integrated with a Si photonic waveguide and operated as a metal-graphene-metal photoconductive detector. Arrayed bowtie-shaped nano-sized metallic structures are integrated to excite SPPs and enhance the external responsivity of the photodetector. The 6 μm long device exhibits a high external responsivity of 0.5 A/W under a -0.4 V bias with a frequency response exceeding 110 GHz. Moreover, operation over a broad spectral range is validated

(across the S, C, and L bands of the optical communication windows). As a result of the high responsivity and large bandwidth of our device, 100 Gbit/s data reception of two-level PAM-2 and four-level PAM-4 encoded signals are demonstrated. To the best of our knowledge, this is the highest responsivity reported so far for waveguide-integrated graphene photodetectors while providing highest speed capabilities for 100 Gbit/s and beyond optical data communication applications. In addition, the fabless Si photonic chip fabrication and the use of chemical vapor deposition (CVD)-grown graphene show the perspective for a large wafer-scale device fabrication and integration.

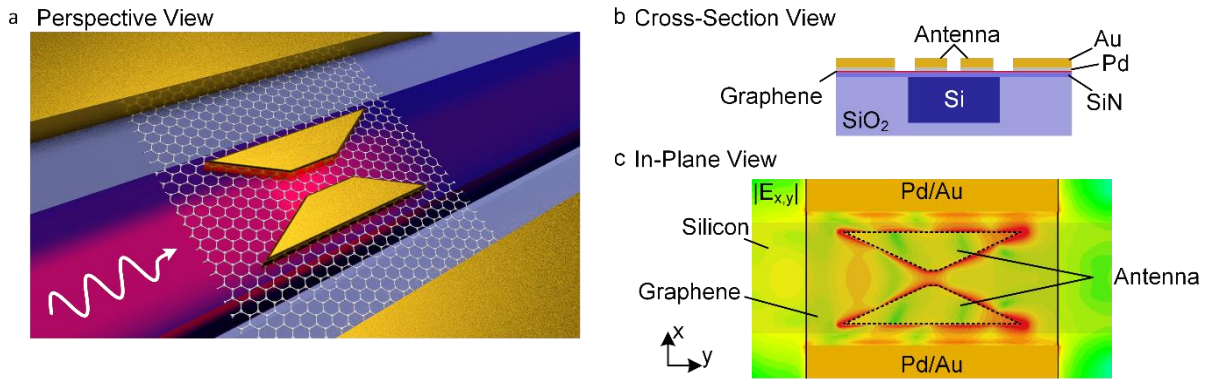

*Fig. 1: Plasmonically enhanced waveguide-integrated graphene photodetector. (a) Three-dimensional (3D) schematic of the photodetector. Light is accessed by a Si waveguide, evanescently coupled to the graphene photodetector which is enhanced by the integrated bowtie-shaped metallic structures. (b) A cross sectional view of the graphene photodetector showing the layer stack. (c) In-plane view of the electric field of the device, obtained by 3D full-wave finite-element method (FEM) simulations.*

**Design, Operation Principle and Fabrication**

Figure 1a and b illustrate the conceptual schematic and the layer stack of the proposed photodetector, respectively. A patterned layer of graphene covers a planarized Si photonics wafer with buried Si waveguides. Metallic bowtie-shaped plasmonic structures with gaps in the nanometer scale are placed on top of graphene and symmetrically aligned with the Si waveguide underneath.

The operation principle is as follows: Linearly TE polarized light is launched through the Si photonic waveguide into the detector. The in-plane electric fields couple evanescently to the graphene photodetector by exciting SPPs in the plasmonic slot of the bowtie-shaped nano-sized metallic structures. The structure is designed to focus the traveling SPPs into the nano-sized gap resulting in gap SPPs with dominant in-plane electric fields. The strongly enhanced in-plane electric field interacts with graphene, which efficiently absorbs in-plane fields[21,44]. The absorbed light in the gap then changes the conductivity in the graphene layer. This change in the conductivity leads to a change in the photoconductive current when a bias voltage is applied across the metallic pads. TM light in the Si waveguide with a field perpendicular to graphene would not really be absorbed and therefore the plasmonic design would not yield an enhancement.

The field enhancement effect in the gap through the plasmonic structure can be best seen with the help of Figure 1c. Figure 1c shows an in-plane view of the electric field of the waveguide-integrated graphene photodetector, simulated by three dimensional (3D) full-wave finite-element method (FEM). The simulation shows how the SPPs in the nano-sized metallic gap structure and around the structure induce strong semi-localized electric fields parallel to the graphene plane and hereby enhance the interaction with graphene. Numerical simulations predict an enhancement of the graphene absorption by a factor of 8.5 for a single field-enhancing element compared to a monolayer graphene of the same length. To further boost the light-matter interaction and increase the graphene absorption one can add more field-enhancing elements, to e.g. an array of 5 elements. Each additional element will contribute less to the overall efficiency with 5 elements being a good number according to our simulations.

The effect of the absorbed light on graphene deserves more discussion. It is known that graphene photodetectors commonly rely on the photovoltaic (PV)[45], the photo-thermoelectric (PTE)[33,45] or the photo-bolometric (PB) effect[46-48]. Under zero bias condition, the bolometric effect does not contribute to the photoresponse, because it can only be observed in biased devices[46]. Under biased conditions, all effects could contribute. If the PV effect were dominant, one would find an increased conductivity with the incident light. However, we detect a decrease in the conductivity with incident light which is indicative for a either the PTE or the PB effect[46]. In our device the graphene layer is doped by the metallic structures[13]. And indeed, in case of strongly doped graphene the PB effect dominates the photodetection[46]. The PB response is induced by photo-generated hot carriers, which modify the channel resistance by either a change to the number of carriers or a change of the temperature dependent carrier mobility[49]. Thus, if a bias voltage is applied between two contacts, the change of graphene resistance can be detected by the change of the current flowing through the metal-graphene-metal photoconductive configuration of our device.

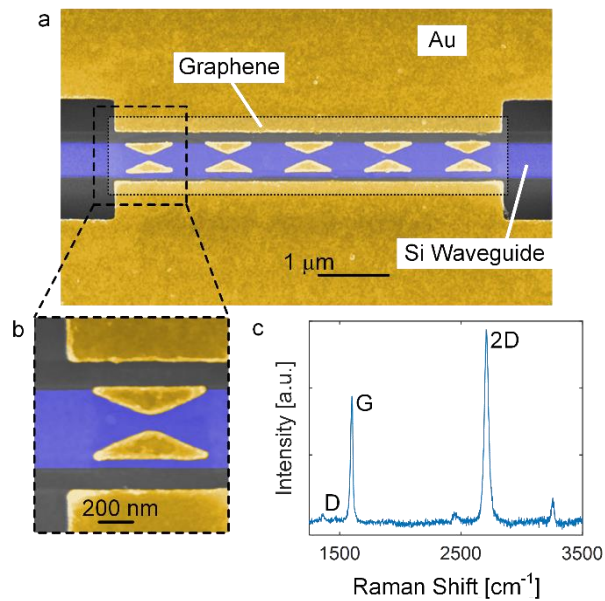

*Fig. 2: Device fabrication.* (a) A top-view false-color scanning electron microscope (SEM) image of a fabricated photodetector, showing the access Si photonic waveguide (in violet color), five arrayed bowtie-shaped plasmonic structures (in yellow color) and contact pads (in yellow color). The dotted line indicates the monolayer graphene under the metallic structures. (b) An enlarged-view false-color SEM image of one field enhancing plasmonic structure with a gap in nanometer scale. (c) Raman spectroscopy of the monolayer graphene transferred on the chip.

For the experimental verification devices have been fabricated on the Si photonic platform. The basic Si photonic structures such as buried waveguides, grating couplers (GCs), and the planarized surface were built by a commercial Si photonic foundry. Before graphene transfer, a 10 nm thick silicon nitride (SiN) layer was first deposited by atomic layer deposition (ALD) to electrically isolate graphene and Si waveguides. The CVD-grown graphene was then transferred to the whole Si photonic chip (≈36 mm$^2$ in size) via a wet transfer technique. Two identical chips with respectively one and double layer graphene were fabricated. Metallic structures including plasmonic field-enhancing structures and contact pads made with 2 nm thick palladium (Pd) and 23 nm thick gold (Au) on top were fabricated by e-beam evaporation and a lift-off process. Pd was used to reduce the contact resistance interfacing with graphene[42,50]. The redundant graphene was finally removed by an oxygen plasma dry etching process. Figure 2a shows a top-view scanning electron microscope (SEM) image of a fabricated device. Five field-enhancing metallic elements in an array were placed and aligned to the Si waveguide as indicated. Figure 2b is an enlarged-view SEM image of one element, showing the nanometer-sized gap. The Raman spectrum as shown in Fig. 2c exhibits typical features of a monolayer graphene[51]. The analysis of the Raman spectrum reveals a full width at half maximum (FWHM) of the 2D peak of 39 cm$^{-1}$ and a peak intensity ratio for the D and G peaks of $I_D/I_G$ = 0.066. This corresponds to an estimated defect density[52,53] in graphene of $1.2 \times 10^{10}$ cm$^{-2}$. A maximum field effect mobility of 1900 cm$^2$/Vs was measured using a reference structure on the same wafer.

**Static Response of the Photo Detectors**

The static characteristics of the photodetector were examined first. Linearly polarized laser light at wavelengths around 1550 nm was coupled on chip via GCs and fed to the photodetector by access Si waveguides. A high precision voltage source was used to apply the bias voltage and measure the current. Photodetectors with five field-enhancing metallic elements were measured under dark and with light illumination at a wavelength of 1550 nm.

To confirm that the dominant effect governing the detector is the photo-bolometric (PB) effect we performed photocurrent measurements. From Fig. 3a at zero bias it can be seen that the overall contribution from the PV effect under illumination with $P_{in}$ =80 µW is small. When a positive voltage bias ($V_{bias}$) is applied the current ($I$) decreases. The current decrease $\Delta I$ gives testimony of a negative conductivity – which is characteristic to the PB effect.

The bias dependent responsivities for single and double-layer graphene devices are plotted in Fig. 3b. For the single-layer graphene detector, the derived external responsivity reaches 0.5 ± 0.05 A/W at a -0.4 V bias and exhibits a quasi-linear dependency. Under low biases the bolometric photocurrent scales with the applied bias according to $I_{ph} = V_{bias}(dG/dT)\Delta T$, where $V_{bias}$ is the bias voltage, $(dG/dT)\Delta T$ is the heat-induced conduction change, and $T$ is the temperature[46]. To get a better insight into the photo-absorption and carrier extraction processes it is of interest to calculate the absorption efficiency $\eta_a$ and internal quantum efficiency (IQE) $\eta_i$. From simulations we derive a plasmonically enhanced graphene absorption of 46% and an IQE of 87%. In good agreement with our measurements this leads to an external responsivity of 0.5 A/W. The double-layer graphene detector supports higher bias voltages. At a -0.6V bias we find a responsivity of 0.4 ± 0.05 A/W. The slope of the plot indicates that higher responsibilities could be obtained at even lower bias voltages. We stopped the characterization at this point in order not to destroy the double-layer detector.

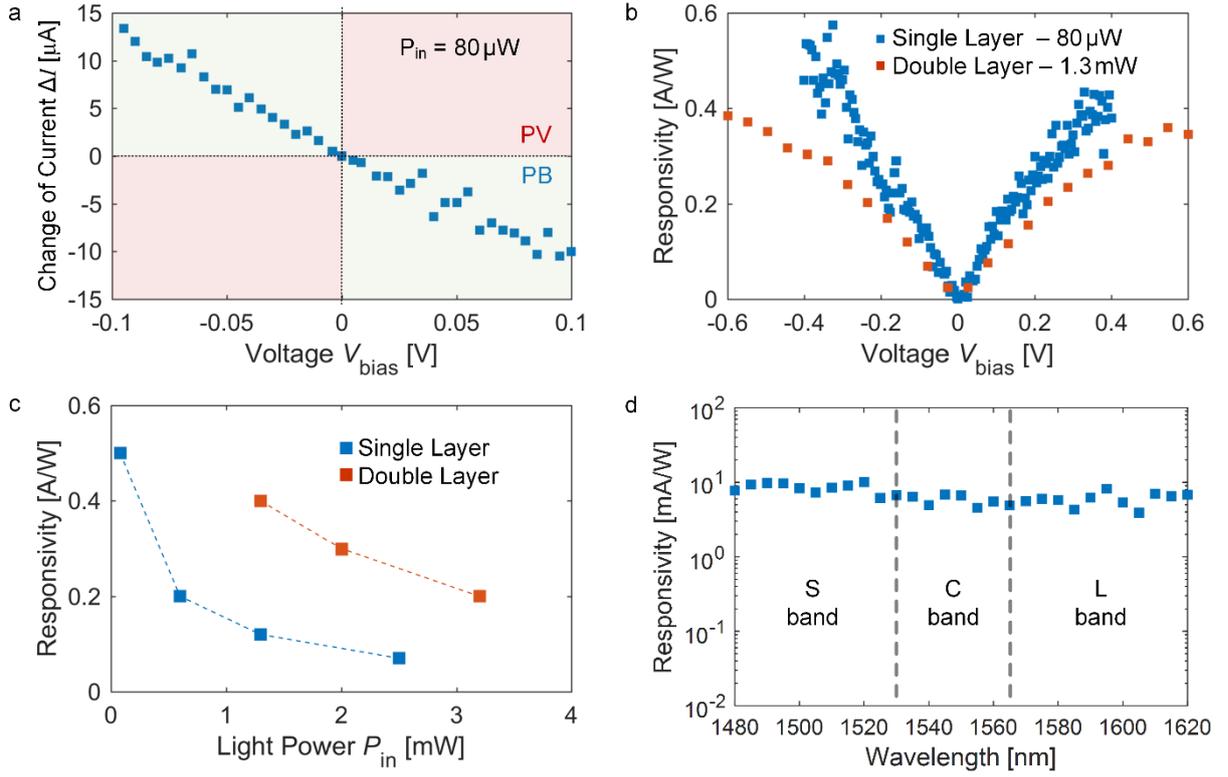

*Fig. 3: Performance characteristics.* (a) Measured change of conductivity as a function of the bias voltage. A negative conduction change for a positive bias indicates a bolometric effect. (b) Responsivity as a function of applied bias voltages for a single-layer (blue scatters) and a double-layer graphene device (red scatters). A photoresponsivity of 0.5 A/W is obtained under a bias of -0.4 V for a single-layer graphene device as measured at input powers of (80 µW). The double-layer graphene device features a higher output saturation power and can achieve 0.4 A/W at a bias of -0.6V for an input signal of 2 mW. (c) Measured responsivity as a function of the light input power. (d) Measured external responsivity as a function of the wavelength of the incident light at zero bias.

The output power of the single-layer graphene device saturates for large input power, resulting in a drop in the external responsivity for input power beyond 100 µW as seen in Fig. 3c. Yet, the double-layer graphene device can handle higher input powers, displaying an external responsivity of 0.4 A/W for a 2 mW input power.

At zero bias, the dark current of the device is as small as 20 nA. Under moderate bias the dark current increases to several hundred microamperes. Yet, the resulting dark current noise power does not lead to a signal-to-noise ratio (SNR) penalty. This is evidenced by the clear open eye diagram from the optical data reception experiment discussed below.

The wavelength-dependent responsivity of the device is checked and shown in Fig. 3d, measured with a tunable laser source. A flat response can be observed, covering from S-band (1480-1565 nm), entire C-band (1530-1565 nm) to L-band (1565-1620 nm) of the optical communication spectral ranges. The measurement range was only limited by the available source and GC spectral response. Simulations show that efficient photodetection can span from at least from 1300 nm to 1650 nm.

**High-Frequency Response and Data Experiments**

To characterize the high-speed performance of the device we measured the frequency response, see Fig. 4. A mostly flat frequency response can be seen spanning the range from 100 kHz up to 110 GHz. Fluctuations in the response are visible but no cut-off trend can be identified. Two distinct setups were employed to cover the broad frequency range. The lower normalized frequency response up to 30 GHz was measured with a continuous wave laser modulated by a commercially available high-speed Mach-Zehnder intensity modulator (MZM, 3dB bandwidth: 30 GHz). At high frequencies, the heterodyne approach[32] was used to generate the radio-frequency (RF) signal from 500 MHz by superposition of two lasers with a varying frequency offset. A second device from the same chip and with the same structure was used for the heterodyne frequency measurement up to 67 GHz, as the first device was damaged during the preparation of the chip for the 67 GHz bandwidth testing. A frequency overlap was intentionally measured to normalize between two measurements. As displayed in Fig. 4, for a device under -0.4 V bias condition a flat frequency response is clearly visible at least to 67 GHz which is the instrumentation-limit of the electrical spectrum analyzer (ESA) used in the measurement. To test the photodetector towards an even higher frequency, we used an ESA offering bandwidth up to 110 GHz. At the higher frequencies we used a double-layer graphene with five metallic elements. The double-layer detector was used as it yields a better SNR and allows measurements with less noise up to highest frequencies. As shown in Fig. 4, the device exhibits a fast photoresponse extended further to 110 GHz. The photo-excited electrons and holes thermalize with other carriers and may cool down via interaction with optical phonons. The relaxation time of these hot carriers in graphene is reported to be a few picoseconds[4]. Eventually, the bandwidth of a PB effect-based photodetector is limited by the product of the electronic heat capacitance and thermal resistance[47]. As for our device, no apparent sign of a cut-off is observed in the bandwidth measurement, indicating the potentials and capabilities of the device operating at even high frequencies.

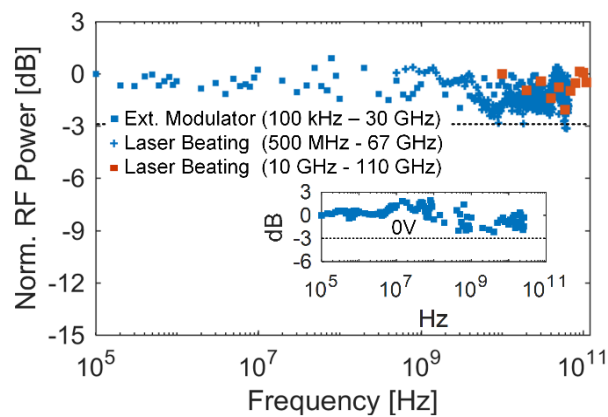

*Fig. 4: Flat frequency response of the graphene photodetector in the range of 100 kHz up to 110 GHz. Blue scatters in the lower frequency range show the frequency response measured with an externally modulated laser; blue crosses and red scatters in the higher frequency range show the response measured with a two-laser beating approach for the single-layer and double-layer graphene devices, respectively. The single and double-layer graphene detectors were biased at -0.4V and -1.0V, respectively.*
*Inset: Frequency response of the device under zero bias condition, (photovoltaic response) measured with an externally modulated laser.*

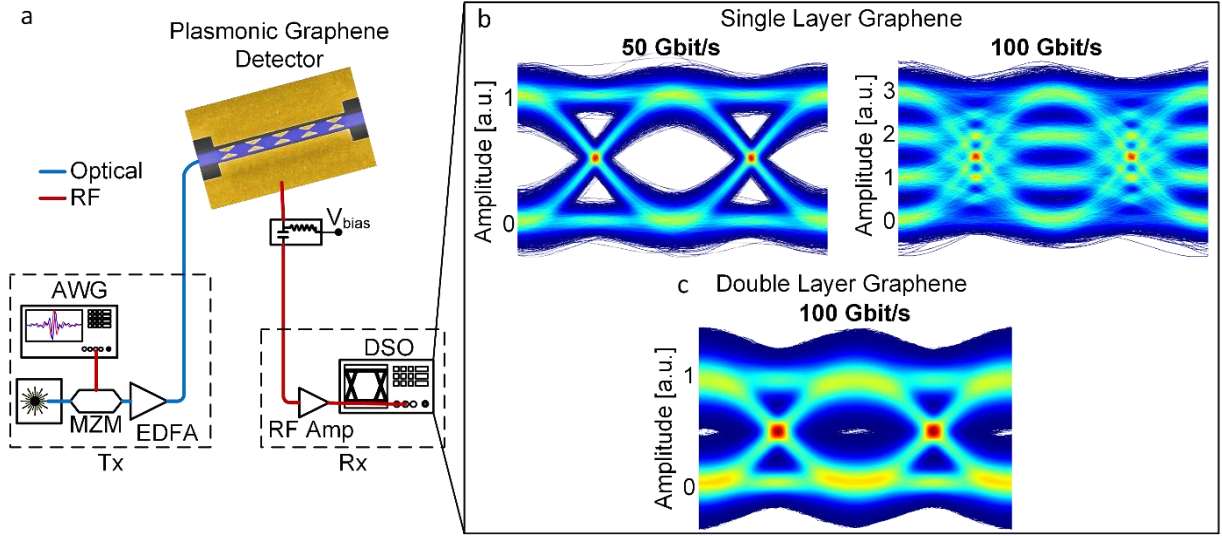

*Fig. 5: Data reception.* (a) Schematic of the measurement setup as used for the data experiments. The setup consists of a continuous wave laser around 1550 nm, an arbitrary waveform generator (AWG), a Mach-Zehnder modulator (MZM), an erbium-doped fiber amplifier (EDFA), a bias-tee, a radio-frequency (RF) power amplifier, and a digital sampling oscilloscope (DSO). Detected eye diagrams of (b) 50 Gbit/s PAM-2 (left), 100 Gbit/s PAM-4 (right) optical signals with a single-layer graphene device, and (c) 100 Gbit/s optical signals with a double-layer graphene device.

The frequency response for device operated at zero bias is shown as inset of Fig. 4. The flat response reveals the corresponding photovoltaic photoresponse is also a fast process, as the drift paths of the photocarriers are very short in the present device.

Data reception experiments were performed to demonstrate the capabilities of the developed devices for 100 Gbit/s high-speed optical data communication applications. The measurement setup is shown in Fig. 5a. The commercially available modulator was used to encode a random bit sequence onto a laser source around 1550 nm. A square-root-raised cosine pulse shaped electrical signal was generated by an arbitrary waveform generator (AWG, sampling rate: 64 GSa/s, electrical 3 dB bandwidth: 25 GHz). The optical signal was amplified with an erbium-doped fiber amplifier (EDFA) and coupled into the single-layer graphene device via GCs and on-chip Si waveguides (~12 dBm). The generated electrical signal was read out with a high-speed ground signal (GS) microwave probe. A bias-tee was used to apply a direct current (DC) voltage of 0.4 V to the graphene detector. The generated RF signal was amplified by a 50 Ω power amplifier and recorded by a real-time digital storage oscilloscope (DSO, sampling rate: 80 GSa/s, electrical 3 dB bandwidth: 33 GHz). Digital signal processing was performed offline to evaluate the bit-error-ratio (BER), including timing recovery, adaptive least mean square (LMS) equalization, non-linear pattern dependent equalization and symbol decision. Figure 5b shows the detected electrical eye diagram for 50 Gbit/s PAM-2 signals below a BER of $2 \times 10^{-4}$ and the detected eye diagram for 50 GBd PAM-4 signals with a line rate of 100 Gbit/s, which is demonstrated for the first time for a graphene photodetector. A BER of $8.9 \times 10^{-3}$ was measured at this data rate. The testable line rate is instrumentally limited, for instance, by the modulator used in the experiment which has a 3 dB bandwidth of 30 GHz.

Reception of a 100 GBd data signal was tested with the double-layer graphene detector. The higher output power provided the necessary signal power to perform experiments at such high speed. For detection the detector was biased with 0.8 V. The power of the signal in the waveguide before the detector was about 11 dBm. For testing a random bit sequence of $2^{18}$ bits was generated by a digital-to-analog converter (DAC) with a sampling rate of 100 GSa/s. Figure 5c shows the eye diagram of the detected 100 Gbit/s PAM-2 signals. A BER of $5.3 \times 10^{-4}$ was found, which is well below the hard-decision forward error correction limit.

**CONCLUSION**

In conclusion, we propose and experimentally demonstrated a compact plasmonically enhanced waveguide-integrated graphene photodetectors by introducing arrayed bowtie-shaped field enhancing plasmonic structures. Devices of 6 μm length feature a high external responsibility of 0.5 A/W and a broad bandwidth exceeding 110 GHz and are capable of detecting optical data at speeds beyond 100 GBd. The detectors have been tested for reception of 100 Gbit/s PAM-2 and 100 Gbit/s PAM-4 modulated optical data, validating the performance and revealing the great potentials of the proposed graphene photodetector next generation optical and microwave photonics communications applications.

**Funding**

The European ERC PLASILOR project (670478) and ETH project grant ETH-45 14-2 are acknowledged for partial funding of the work.

**Acknowledgements**

This work was carried out partially at the Binning and Rohrer Nanotechnology Center (BRNC) and in the FIRST lab cleanroom facility at ETH Zurich. We are grateful to H. R. Benedickter and A. Olziersky for the help with the measurement and fabrication, respectively.

†These authors contributed equally to this work.